Proceedings of
# VIII International Workshop on
# Locational Analysis and
# Related Problems

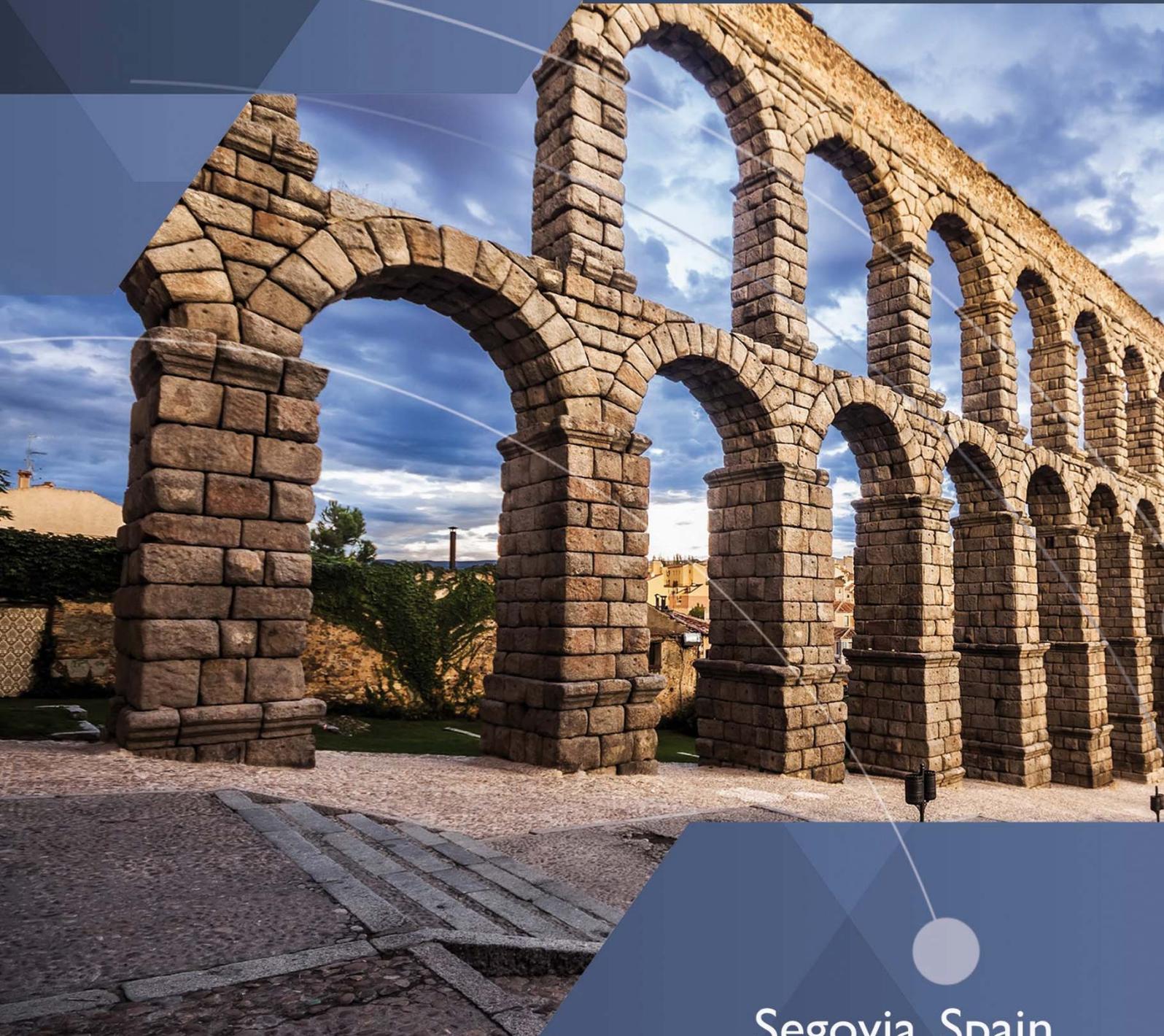

Segovia, Spain
September 27-29, 2017

Program Overview

| | Wednesday Sep. 27th | Thursday Sep. 28th | Friday Sep. 29th |
|---|---|---|---|
| 9:00-10:40 | | SESSION 2: CONTINUOUS LOCATION | SESSION 6: NETWORK/TERRITORY PLANNING |
| 10:40-11:10 | | Coffee break | Coffee break |
| 11:10-12:30 | | Invited Speaker: Pierre Bonami | Invited Speaker: James Campbell |
| 12:30-13:45 | | SESSION 3: ROUTING PROBLEMS | SESSION 7: HUB LOCATION |
| 13:45-14:30 | | LUNCH | Locat. Network Meeting |
| 14:30-15:30 | | | LUNCH |
| 15:30-17:10 | | SESSION 4: APPLICATIONS | |
| 17:10-17:40 | REGISTRATION | Coffee break | |
| 17:40-18:10 | OPENING SESSION | SESSION 5: DISCRETE LOCATION | |
| 18:10-19:20 | SESSION 1: DISCRETE LOCATION | | |
| 19:20-19:50 | | | |
| 21:00 | Welcome Reception | DINNER | |

# PROCEEDINGS OF
# THE VIII INTERNATIONAL WORKSHOP
# ON LOCATIONAL ANALYSIS AND
# RELATED PROBLEMS (2017)

Edited by

Marta Baldomero-Naranjo

Inmaculada Espejo-Miranda

Luisa I. Martínez-Merino
Antonio M. Rodríguez-Chía

Diego Ruiz-Hernández

# Preface

The International Workshop on Locational Analysis and Related Problems will take place during September 27-29, 2017 in Segovia (Spain). It is organized by the Spanish Location Network and Location Group GELOCA (SEIO). GELOCA is a working group on location belonging to the Statistics and Operations Research Spanish Society. The Spanish Location Network is a group of more than 100 researchers distributed into 16 nodes corresponding to several Spanish universities. The Network has been funded by the Spanish Government.

Every year, the Network organizes a meeting to promote the communication between its members and between them and other researchers, and to contribute to the development of the location field and related problems. Previous meetings took place in Málaga (September 14-16, 2016), Barcelona (November 25-28, 2015), Sevilla (October 1-3, 2014), Torremolinos (Málaga, June 19-21, 2013), Granada (May 10-12, 2012), Las Palmas de Gran Canaria (February 2-5, 2011) and Sevilla (February 1-3, 2010).

The topics of interest are location analysis and related problems. It includes location, networks, transportation, routing, logistics models, as well as, exact and heuristic solution methods, and computational geometry, among others.

The organizing committee.

## Scientific committee:

- Emilio Carrizosa (U. de Sevilla)
- Ángel Corberán (U. Valencia)
- Elena Fernández (U. Politécnica de Cataluña)
- Alfredo Marín (U. de Murcia)
- Juan A. Mesa (U. de Sevilla)
- Blas Pelegrín Pelegrín (U. de Murcia)
- Justo Puerto (U. de Sevilla, España)
- Antonio M. Rodríguez Chía (U. de Cádiz)
- Diego Ruiz Hernández (Colegio Universitario de Estudios Financieros)

## Organizing committee:

- Marta Baldomero Naranjo (U. de Cádiz)
- Inmaculada Espejo Miranda (U. de Cádiz)
- Luisa Isabel Martínez Merino (U. de Cádiz)
- Belén Palop del Río (U. de Valladolid)
- Dolores Rosa Santos Peñate (U. de Las Palmas de Gran Canaria)

# Contents

PROGRAM

# Wednesday September 27th

## 17:10-17:40 Registration

## 17:40-18:10 Opening Session

## 18:10-19:50 Session 1: Discrete Location

A bilevel approach for the single-source capacitated facility location problem with customer's preferences
H.I. Calvete, C. Galé, J.F. Camacho-Vallejo, and M.S. Casas-Ramírez

Approval voting problem under the $k$-centrum criterion
D. Ponce, J. Puerto, F. Ricca, and A. Scozzari

A stochastic multi-period covering model
A. Marín, L.I. Martínez-Merino, A. M. Rodríguez-Chía, and F. Saldanha-da-Gama

The mobile facility location problem
M. Landete

## 21:00 Welcome Reception

# Thursday September 28th

## 9:00-10:40 Session 2: Continuous Location

Some criteria for locating sensors in a wind turbine blade
M.C. López-de-los-Mozos, J.A. Mesa, D. Ruiz-Hernández, and C.Q. Gómez-Muñoz

Territorial districting models for the reorganization of postal services
G. Bruno, M. Cavola, A. Diglio, and C. Piccolo

On the location of separating hyperplanes with $\ell_p$-norms margins
V. Blanco, J. Puerto, and A.M. Rodríguez-Chía

Location theory and some physical principles in a nutshell
J. Puerto

## 10:40-11:10 Coffee break

## 11:10-12:30 Invited Speaker: Pierre Bonami

Some recent advances in mixed-integer nonlinear optimization
P. Bonami

## 12:30-13:45 Session 3: Routing

Routing vehicle fleets during disaster relief
E. Barrena, D. Canca, and F.A. Ortega

The periodic rural postman problem with irregular services
E. Benavent, Á. Corberán, D. Laganà, and F. Vocaturo

The periodic vehicle routing problem with driver consistency
I. Rodríguez-Martín, J.J. Salazar-González, and H. Yaman

## 13:45-15:30 Lunch

## 15:30-17:10 Session 4: Applications

Energy-efficient timetables
D.C. Ortíz, and A. Zarzo

On location and vessel fleet composition for offshore wind farm maintenance

A. Gutierrez, E.M.T. Hendrix, G. Ortega, D. Haugland, E.E. Halvorsen-Weare

Blackout risk mitigation by using distributed gas turbine generation. An application to the electrical Spanish distribution network.
D. Canca, Á. Arcos-Vargas, and F. Nuñez

Inducing universal access to privately-managed social-interest goods via location decisions
J. Elizalde, A. Erro, and D. Ruiz-Hernández

## 17:10-17:40 Coffee break

## 17:40-19:20 Session 5: Discrete Location

Profiling the inherent complexity of different facility location strategies
J.M. Pinar-Pérez, D. Ruiz-Hernández, and M. Menezes

Uncertainty in building times: identifying critical facilities in a dynamic location problem
J. Dias

Some heuristic methods for the $p$-median problem with maximum distance constraints
A. Esteban Pérez, and J. Sáez-Aguado

Facilities delocation in the retail sector
M. Sierra-Paradinas, A. Alonso-Ayuso, and J.F. Rodríguez-Calo

## 21:00 Dinner

# Friday September 29th

## 9:00-10:40 Session 6: Network/Territory design

A matheuristic for the rapid transit network design problem with elastic demand
D. Canca, A. De-Los-Santos, G. Laporte, and J.A. Mesa

Supply chain complexity and the network design: Location does matter!
M.B.C. Menezes, and D. Ruiz-Hernández

The effect of products' short lifecycle on network design
D. Ruiz-Hernández, M.B.C. Menezes, and O. Allal-Cheriff

The multi-Period service territory design problem
M. Bender, J. Kalcsics, A. Meyer, S. Nickel, and M. Pouls

## 10:40-11:10 Coffee Break

## 11:10-12:30 Invited Speaker: James F. Campbell

Strategic design of drone delivery systems
J.F. Campbell

## 12:30-13:45 Session 7: Hub Location

Tree of hubs location problem with upgrading
A. Marín

The ordered median tree of Hubs location problem
M.A. Pozo, J. Puerto, and A.M. Rodríguez-Chía

Heuristics for the stochastic uncapacitated $r$-allocation $p$-hub median problem
J. Peiró, Á. Corberán, R. Martí, and F. Sandanha-da-Gama

## 13:45-14:30 Location Network Meeting

## 14:30-15:30 Lunch

INVITED SPEAKERS

# Some recent advances in mixed-integer nonlinear optimization

Pierre Bonami[1]

[1]*Laboratoire d'Informatique Fondamentale de Marseille Aix Marseille Université/ IBM, France,*  pierre.bonami@es.ibm.com

Mixed Integer Nonlinear Optimization is the optimization of a nonlinear function over a feasible set described by nonlinear functions and integrality constraints. We will review some of the main algorithmic techniques that are employed in commercial solvers. We will focus in particular on two recent works that address global solution of non-convex mixed integer optimization problems: cuts from the binary quadric polytope and maxclique inequalities.

# Strategic design of drone delivery systems

James F. Campbell[1]

[1]*College of Business Administration, University of Missouri – St. Louis, USA,*
campbell@umsl.edu

Home delivery by drones as an alternative to traditional delivery by trucks is attracting considerable attention from major retailers and service providers (Amazon, UPS, Google, DHL, Wal-mart, etc.), as well as from startups. While drone delivery may offer considerable economic savings, the fundamental issue of how best to deploy drones for home delivery is not well understood. Operations Research has a long tradition in analyzing location, logistic and routing problems and drone delivery provides some new opportunities for research. This presentation first provides an overview of drone delivery systems and highlights research opportunities, especially using drones is conjunction with trucks. Then we present a strategic analysis for the design of truck-drone delivery systems using continuous approximation modeling techniques to derive general insights. We formulate and optimize models that consider hybrid truck-drone delivery (where truck-based drones make deliveries simultaneously with trucks), truck-only delivery and drone delivery from depots. Results show that truck-drone hybrid delivery can be very economically advantageous in many settings, especially in more rural areas and with multiple drones per truck, but that the benefits depend strongly on the relative operating costs and marginal stop costs. Results also examine locating depots for drones, especially to achieve high service levels.

ABSTRACTS

# Routing vehicle fleets during disaster relief[*]

Eva Barrena,[1] David Canca,[2] and Francisco A. Ortega[3]

[1]*Universidad de Granada,* ebarrena@ugr.es

[2]*Universidad de Sevilla,* dco@us.es

[3]*Universidad de Sevilla,* riejos@us.es

Last mile in the humanitarian logistics management refers to the supply of relief items from the local distribution centres to the disaster area. Some estimations suggest that as much as 80% of the expenditure of aid agencies lies on logistics. Therefore, humanitarian logistics management need to be efficiently and effectively planned in their relevant components (facility location, inventory management, transportation management, distribution management, etc.) at the strategic, tactical and operational levels. The problem addressed in this paper consists of designing routes for vehicles among nodes that receive an available quantity of goods (depots called local distribution centers –LDCs-) and others that have a demand of those goods, simultaneously choosing the most adequate types of vehicles and determining the flow of the aid along a time horizon. For that reason, we consider the possibility of interaction between the different LCDs in terms of transferring goods to be supplied and allocating vehicles for transportation.

[*]This research has been partially supported by the Spanish Ministry of Economy and Competitiveness through grant MTM2015-67706-P (MINECO/FEDER, UE). This support is gratefully acknowledged.

# The periodic rural postman problem with irregular services

Enrique Benavent,[1] Ángel Corberán,[1] Demetrio Laganà, [2] and Francesca Vocaturo[2]

[1]*DEIO, Universitat de València, Dr. Moliner 50, 46100 Burjassot, Valencia, Spain,*   enrique.benavent@uv.es
angel.corberan@uv.es

[2]*DIMEG, Università della Calabria, via Pietro Bucci Cubo 41/C, 87036 Arcavacata di Rende (CS), Italy,*   demetrio.lagana@unical.it
francesca.vocaturo@unical.it

Periodic routing problems consist of designing vehicle routes for all the days of a given time horizon, or planning period, in order to meet specific service requirements of a subset of arcs in a given graph, which usually represents a street or road network. Generally, a required arc does not need a service on every day, but must be serviced at least once (or a specified number of times) over the time horizon. In some practical applications the required frequency of services may be irregular; for instance, some arcs must be serviced twice during the first five days of a week and once during the weekend; in addition, the days for service may vary from one week to the other.

In this talk, we deal with the periodic rural postman problem with irregular services (PRPPIS) in which some links of a mixed graph must be traversed a specified number of times in some given time horizon sub–periods. The aim is to design a set of least-cost tours, one for each period in the horizon, that satisfy the service requirements. Some practical applications of the problem can be found in road maintenance operations and road network surveillance.

In order to solve the PRPPIS, we propose a mathematical model and a branch-and-cut algorithm. In the solution framework, constraints ensuring connectivity and other valid inequalities are identified by using specific separation procedures. Some valid inequalities consider the particular nature of the PRPPIS. We show the effectiveness of the solution approach through an extensive experimental phase.

# The multi-period service territory design problem

Matthias Bender,[1] Jörg Kalcsics,[2] Anne Meyer,[1] Stefan Nickel,[1,3] and Martin Pouls[1]

[1]*Department of Logistics and Supply Chain Optimization, Research Center for Information Technology (FZI), Karlsruhe, Germany*

[2]*School of Mathematics, University of Edinburgh, Edinburgh, Scotland*

[3]*Institute of Operations Research, Karlsruhe Institute of Technology (KIT), Karlsruhe, Germany,*  stefan.nickel@kit.edu

Classical sales or service territory design problems consist of grouping customers into larger clusters, which are called territories or districts, such that some relevant planning criteria, e.g., compactness and balance, are met [2]. In each district, a service provider, e.g., a salesperson or service technician, is responsible for providing services at the customers' sites. In many cases, these services must be provided several times during a given planning horizon, which extends the classical problem to a multi-period setting.

The main contributions of this talk are the following:

- We introduce a new problem, which we call the Multi-Period Service Territory Design Problem (MPSTDP) ( [1] ). Despite its high practical relevance, it has not been studied in the literature before.

- We formally define the scheduling subproblem, i.e., the subproblem dealing with the assignment of service visits to the weeks and days of the planning horizon, as a mixed integer linear programming model.

- We propose a heuristic solution approach for the scheduling subproblem. The approach is capable of considering the relevant planning requirements of practical settings. It involves the repeated solution of an integer programming model, which can easily be extended by additional planning requirements.

- We present a column generation formulation of the resulting scheduling problem and propose an exact branch-and-price algorithm. Our method incorporates specialized acceleration techniques, such as a fast pricing heuristic and a symmetry handling strategy. The latter aims to reduce the symmetry inherent to the model by fixing variables and thereby eliminating symmetric solutions from the search tree.

- We perform extensive computational experiments on real-world instances and on instances that were derived from real-world data by varying the values of some parameters. The results show that the new approach produces high-quality solutions and outperforms solution methods of existing software products.

# On the location of separating hyperplanes with $\ell_p$-norms margins

Víctor Blanco,[1] Justo Puerto,[2] and Antonio M. Rodríguez-Chía[3]

[1]*Universidad de Granada,*  vblanco@ugr.es

[2]*Universidad de Sevilla,*  puerto@us.es

[3]*Universidad de Cádiz,*  antonio.rodriguezchia@uca.es

In this work we present new results on the determination of a hyperplane that separate two classes of given data. We deal with the problem of obtaining such a hyperplane when the goal is to minimize the $\ell_p$-margin (with rational $p \geq 1$) between the two classes. We provide valid non-linear mathematical programming formulations for the problem involving the minimization of homogenous polynomials over linear regions. The formulation allows us to manage the use of transformations of the data and the resolution of the problem without using the specific knowledge of the transformation but some properties about it. Some methodologies for solving the problem in practice are provided.

## Introduction

We are given a set of $d$ quantitative measures about $n$ individuals. The $d$ measures about each individual $i \in \{1, \ldots, n\}$ are identified with the vector $\mathrm{x}_{i\cdot} \in \mathbb{R}^d$. Each observation is also classified into a class in $\{-1, 1\}$, being $y_i \in \{-1, 1\}$ the class of the $i$th observation, for $i = 1, \ldots, n$.

The goal of this work is to find a hyperplane $\mathcal{H} = \{z \in \mathbb{R}^d : \omega^t z + b = 0\}$ that minimizes the misclassification of data to their own class.

By using a classical result in [2], the problem can be formulated as:

$$\rho^* = \min\|\omega\|_q^q + C\sum_{i=1}^{n}\xi_i$$
$$\text{s.t. } y_i(\omega^t\mathrm{x}_{i\cdot} + b) \geq 1 - \xi_i, \qquad \forall i = 1,\ldots,n,$$
$$\xi_i \geq 0, \qquad \forall i = 1,\ldots,n.$$

where $q$ is such that $\frac{1}{p} + \frac{1}{q} = 1$ and $C$ is a constant. This problem allows an equivalent reformulation as a second-order cone programming problem.

By using Lagrangian duality, we prove that the above problem can be equivalently reformulated as a set of nonlinear optimization problems where only multivariate homogeneous polynomials of certain degree $r$ are involved.

In case we consider a transformation, $\Phi : \mathbb{R}^d \to \mathbb{R}^D$, which maps the data to a higher dimensional space, better separation schemes can be derived. We prove that applying the same methodologies, we can obtain minimal separating hyperplanes without the explicit *knowledge* of $\Phi$. We determine sufficient conditions on $\Phi$ necessary for obtaining the separating hyperplanes and derive a methodology to classify outsample data. These characterizations involve the use of tools borrowed from real higher-dimensional tensor rank-one decomposition [1].

Finally, we detail how to solve the dual separating hyperplane problem using the Theory of Moments and SDP Relaxations [3].

# Territorial districting models
# for the reorganization of postal services

Giuseppe Bruno,[1] Manuel Cavola,[1] Antonio Diglio[1] and Carmela Piccolo[1]

[1]*Department of Industrial Engineering, University of Naples Federico II*
*Piazzale Tecchio, 80 – 80125, Naples, Italy*
(giuseppe.bruno, antonio.diglio,carmela.piccolo)@unina.it,manuelcavola@live.com

In the last years, postal services were interested by profound changes due to different factors. In particular, the progressive substitution of conventional letters by electronic forms of communication (e-substitution), led to a drastic decrease of volumes of classic postal mails and to a consequent reduction of revenues [3]. In order to effectively face these changes, postal companies are responding by adapting their logistic systems. In this context, we address several problems related to the reorganization of mail collection and delivery service, in collaboration with the Italian Postal Service Provider (Poste Italiane S.p.A.).

Usually, two types of mails may be distinguished: *priority* and *ordinary mails*, that have to be delivered within one and two days respectively. For ordinary mails, mailboxes represent the access points of users to the logistic network. Mails contained in the mailboxes are picked up by dedicated operators, moved to Distribution Centers (DC) for sorting operations, and transported to final destinations. Usually the time span devoted to collection is very limited, as all the subsequent activities (sorting, distribution, delivery) need to be performed within strict deadlines. Then, the number of employed operators and routes strongly depends on the number and on the position of mailboxes over the study region. As the mails volumes have drastically reduced, Poste Italiane is interested in reducing the number of mailboxes, in order to improve service efficiency. However, as the postal service is labeled as "essential", in pursuing this objective it is obliged to satisfy contraints imposed by specific authorities on users accessibility.

In order to tackle the problem, we refer to the class of *Districting Problems*

(DP), that aim at partitioning a given territory in a fixed number of sub-areas, named *districts* [7]. The reference territory is usually divided into basic units, that have to be grouped so as constraints on dimension and topology of single districts are satisfied. DPs are suitable to describe problems related to the organization of public services, in which the goal is to design sub-areas for facilities' service provision [5, 6]. Within such literature stream, some contributions deal with the problem of modifying existing district maps, as a consequence of facility closure or relocation (redistricting problems) [2, 4].

In the case under analysis, the reference territory is divided into census tracks, already grouped in districts, each associated to a single mailbox. In such partition, it has been reasonably assumed that users refer to the closest postal box. Once a subset of mailboxes are closed, users have to be reassigned to the closest active mailbox in the neighborhood of their residence. The problem consists in deciding how many and which mailboxes have to be closed in order to minimize the management costs of the service and guaranteeing a good and equitable service level to users [1].

We tested the proposed models on the case of the urban area of the city of Bologna (Italy), where 272 mailboxes are located to serve about 400.000 inhabitants. Obtained solutions provide useful indications for supporting decision makers in the rationalization of the described process.

# A bilevel approach for the single-source capacitated facility location problem with customer's preferences [*]

Herminia I. Calvete, and Carmen Galé,[1]
José-Fernando Camacho-Vallejo, and Martha-Selene Casas-Ramírez[2]

[1]*Universidad de Zaragoza, Spain,*
herminia@unizar.es, cgale@unizar.es

[2]*Universidad Autónoma de Nuevo León, México,*
jose.camachovl@uanl.edu.mx, martha.casasrm@uanl.edu.mx

Location models are among the main optimization models in facility planning. At the strategic level of decision making, the decisions remain unchanged for a long time. So, to decide where to locate the facilities is one of the most critical decisions in logistics. The facility location problem is a well-known combinatorial optimization problem. It consists of selecting the location of a set of facilities from a finite set of potential sites to meet customer demand. The location problem considered in this work is single source model, that is to say, each customer is served from a single facility. Moreover, a capacity constraint is added to the model to limit the number of customers which can be allocated to each facility. Finally, other crucial aspect to be considered is the choice of the optimization criterium for allocating customers to facilities.

In the literature, the customer allocation problem has been solved considering alternative criteria in the objective function according to different perspectives ( [1], [2], [3] and [4]). In this paper, the customers' preferences regarding the selection of facilities are taken into account. Once the facilities are located, the customers are allocated to their preferred facility while holding the capacity constraints. In order to solve the conflict that arises when several customers prefer the same open facility and its capacity is insufficient to serve all of them, we assume a cooperative behavior of the customers.

[*]This research work has been funded by the Spanish Ministry of Economy, Industry and Competitiveness under grant ECO2016-76567-C4-3-R.

A bilevel problem is proposed to model this system. In the upper level the leader decides on the facilities which will be located. In the lower level the allocation problem is solved. The leader's objective function consists of minimizing the overall costs for opening the facilities and serving the customers. Concerning the lower level problem, we establish a set of ordered preference values for every customer in which the lowest value is associated with the most preferred facility. Then, the optimization criteria of the lower level problem is to minimize the sum of the preferences. Moreover, we assume an optimistic approach when solving the bilevel problem, i.e. in case of multiple optima in the lower level problem the best option for the leader is selected.

An analysis exploiting some particularities of the problem leads to a reformulation as a single level optimization problem. Small instances of this model can be solved using usual optimization tools. A genetic algorithm is developed for solving instances which cannot be solved to optimality in a reasonable computational time. A numerical experimentation is conducted to test the performance of the heuristic algorithm. Benchmark instances by Holmberg for the single source capacity facility location problem are considered and suited for the problem introduced in this work. The number of customers in these instances ranges from 20 to 90 and the number of potential locations ranges from 10 to 30. A new set with larger instances has also been randomly generated.

# Blackout risk mitigation by using distributed gas turbine generation. An application to the electrical Spanish distribution network.[*]

David Canca,[1] Ángel Arcos-Vargas,[2] and Fernando Nuñez[3]

[1]*Universidad de Sevilla,* dco@us.es

[2]*Universidad de Sevilla,* aarcos@us.es

[3]*Universidad de Sevilla,* fnunuezh@us.es

The aim of this paper consists on analysing the economic aspects of the mitigation of network power at risk by locating the appropriate gas turbines at risk points. We propose two different Mixed Integer Programming optimization models. The first one considers the existence of a global network agreement among generation companies and distributor in order to cover at least partially all the points at risk. The second one considers an unregulated framework where distributed generators have full freedom to choose locations among those proposed by the distribution company and to select the most convenient turbine models. In both cases the models consider the temporal deployment of risk mitigation. Since electricity distribution is a regulated activity, we study the impact of a possible regulated remuneration that encourage generators to install the appropriate generation power at the specific points at risk. We illustrate the proposed approach by solving the case of power at risk in a big scenario concerning approximately the half of the Spanish distribution network.

[*]This research has been partially supported by the Spanish Ministry of Economy and Competitiveness through grant MTM2015-67706-P (MINECO/FEDER, UE). This support is gratefully acknowledged.

# Energy-efficient timetables.[*]

David C. Ortíz,[1] and Alejandro Zarzo[2]

[1]*Departamento de Organización Industrial y Gestión de Empresas I (Industrial Engineering and Management Science), Universidad de Sevilla, Spain,*  dco@us.es

[2]*Departamento de Matemáticas del Área Industrial, E.T.S. Ingenieros Industriales, Universidad Politécnica de Madrid, Spain, and Instituto Carlos I, Universidad de Granada, Granada, Spain,*  alejandro.zarzo@upm.es

A methodology to design timetables with minimum energy consumption in Rapid Railway Transit Networks is presented. Using an empirical description of the train energy consumption as a function of running times, the timetable design problem is modelled as a Mixed Integer Non-Linear optimization problem (MINLP) for a complete two-way line. In doing so, all the services in both directions along a certain planning horizon are considered while attending a known passengers' demand. The MINLP formulation, which depends on train loads, is fully linearised supposing train loads are fixed. A sequential Mixed Integer Linear (MILP) solving procedure is then used to solve the timetabling optimization problem with unknown train loads. The proposed methodology emphasizes the need of considering all the services running during the planning horizon when designing energy-efficient timetables, as consequence of the relationship among train speeds, frequency and fleet size of each line. Moreover, the convenience of considering the energy consumption as part of a broad objective function that includes other relevant costs is pointed out. Otherwise, passengers and operators could face up to an increase in the whole cost and a decrease in the quality of service. A real data scenario, based on the C-2

---

[*]This research work was supported by the Ministry of Economy and Competitiveness of Spain and the European Regional Development Fund (ERDF) under grant MTM2015-67706-P. The second author (AZ) also acknowledge partial financial support from the Ministry of Economy and Competitiveness of Spain and the European Regional Development Fund (ERDF) under grant MTM2014-53963-P, from Junta de Andalucía through the Excellence Grant P11-FQM-7276 and the research group FQM-020, and from Technical University of Madrid (research group TACA).

Line of the Madrid Metropolitan Railways, is used to illustrate the proposed methodology and to discuss the differences between the minimum-energy solutions and those obtained when considering operation and acquisition costs.

# A matheuristic for the rapid transit network design problem with elastic demand

David Canca,[1] Alicia De-Los-Santos,[2], Gilbert Laporte[3] and Juan A. Mesa,[4]

[1]*Department of Industrial Engineering and Management Science, University of Seville, Spain,*  dco@us.es

[2]*Department of Statistic, Econometrics, I.O. And Business Organization, University of Cordoba, Spain,*  aliciasantos@uco.es

[3]*Canada Research Chair in Distribution Management. HEC Montreal, Canada.*  gilbert.laporte@cirrelt.ca

[4]*Department of Applied Mathematics II, University of Seville, Spain.*  jmesa@us.es

In this work, we propose a matheuristic for the integrated Railway Rapid Transit Network Design and Line Planning problem. The network design problem incorporates costs relative to the network construction and proposes a set of candidate lines whereas the line planning problem determines the best combination of frequencies and train capacities for the set of lines taking into account rolling stock, personnel and fleet acquisition costs. We consider the existence of an alternative transportation mode competing with the railway system for each origin-destination pair. Passengers choose their transportation mode according to their own utility. Due to the problem complexity and the impossibility of solving the problem on realistic size scenarios, we develop a matheuristic combining an Adaptive Large Neighborhood Search (ALNS) algorithm and an transit assignment model. At each iteration, in an cooperative way, the ALNS solves the network design problem and the assignment model is in charge of the line planning problem. As consequence of the non-linear nature of the assignment problem, in order to guarantee optimality at each iteration, a full-linearisation of the inner model is also presented.

# Uncertainty in building times: identifying critical facilities in a dynamic location problem

Joana Dias[2]

[1]*INESCC, CeBER, Faculty of Economics, University of Coimbra, Av. Dias da Silva, 165, 3004512 Coimbra, Portugal*   joana@fe.uc.pt

In dynamic location problems, most of the times, the decisions that are considered in the mathematical models are where and when to locate facilities. It is assumed that deciding when to open will deterministically determine when to begin all the necessary procedures to guarantee that the facility is indeed opened at the desired time period. These procedures can include property acquisition, infrastructure construction, having human and material resources prepared, among other things. In this work, uncertainty associated with this "building time" is explicitly assumed. The decisions are now where to locate facilities and when to begin building these facilities. As time and cost compromises are usually present, the decision maker can decide to increase costs in order to decrease the uncertainty associated with the building times. It is possible to identify critical facilities: the ones in which it is worth investing to guarantee that there are no delays in their opening times.

**Keywords:**   dynamic location, uncertainty, scenarios, building time, critical facilities

## 1.     Introduction

Most of the times, there are several activities that have to be executed before a facility is ready to be used. There are situations where the facility has to be built, infrastructures have to be created and resources have to be assigned. Opening a facility is usually a project that has to be planned, executed, and that takes time. More often than not, there can be delays. Most of the times, there is uncertainty related with the time lag between beginning to prepare a facility to be open and having it ready to have assigned clients. In the present work, a dynamic location problem will be described, where the uncertainty associated with the time to "build" a facility is explicitly considered. Uncertainty is represented by resorting to scenarios. It is assumed that the location decisions are taken at the beginning of the planning horizon and cannot be changed from that point forward. The assignment variables can be optimized in each time period. Moreover, the decision maker has the possibility of increasing the cost associated with the opening of a given facility in order to decrease the uncertainty related with its building time (by assigning more resources to its preparedness process, for instance).

## 2. Mathematical Model

The most obvious way of formulating this problem is by using a non-linear formulation, where the building times are decision variables themselves that will determine the possible values of the assignment variables. It is, however, possible to also devise a mixed integer linear programming problem resorting to additional auxiliary variables and constraints. It is even possible to consider different levels of investment, with different consequences in the corresponding building times, representing the compromises that exist between building time and costs. Facilities which are advantageous to invest in order to decrease/eliminate the uncertainty associated with their building times can be considered as critical facilities, in the sense that it is worth spending more to guarantee that they are operational in the expected time period.

## 3. Conclusion

In this problem, the uncertainty associated with the building time of facilities in a dynamic location problem is considered. It is possible to develop a mixed integer linear programming problem that represents this problem. Computational experiments are being done, with the purpose of: 1. Understanding how the model can be used to illustrate the time vs cost compromises and its relation with the identification of critical facilities; 2. To assess the possibility of solving large instances of this problem by using a general solver, and to reach a conclusion regarding the need to develop a (meta)heuristic procedure. Different objective functions could be considered, like minimizing the maximum regret, or multiobjective approaches could be devised.

# Inducing universal access to privately-managed social-interest goods via location decisions

Javier Elizalde, [1] Amaya Erro, [2] Diego Ruiz-Hernández [3,4]

[3]*University College for Financial Studies, Department of Quantitative Methods, Leonardo Prieto Castro 2, 28040, Madrid, Spain*   d.ruiz@cunef.edu

[1]*Universidad de Navarra, Campus Arrosadía, 31006, Pamplona, Spain*

[2]*Universidad Pública de Navarra, Campus Universitario, Edificio Amigos, 31009, Pamplona, Spain*

[4]*Supply Chain and Complexity Lab, KEDGE Business School*

There exist a largen number of services for which the public authority may have an interest in guaranteeing universal access. Examples, among many others, are health and education services and the delivery of water, electricity and postal services to the households. Some of these services, such as water, electricity and post, are most often natural monopolies, with one single firm in the most efficient market structure. Moreover, those services are directly delivered to the final consumer and therefore the universal provision of the service is independent –from the point of view of the final consumer– of the location of the sources.

However, in other cases, such as education, health, and community services, ease of access plays a central role in the universal provision of the service. Several approaches have been taken to analyse the problem of finding the optimal location of a service in order to guarantee total population coverage; with efforts coming from disciplines as diverse as economics, geography and operational research. Among the recent articles dealing with the location of education centres we can refer the work of Ewing et al. [2] and Pal [5]; regarding the location of health and emergency services we can cite Daskin and Dean [1]; and Günes and Nickel [3].

In the present work, given that some of these services are provided in facilities where the individuals have to commute to, we use a theoretical model of spatial monopoly. A choice that, moreover, allows us to better illustrate the problem of partial vs full provision. The firm providing the service is assumed to private, emphasising the need of public intervention for guaranteeing universal access. We focus on two public policies that can be used by the government in order to influence the level of access and prices, namely, fixing a universal price with free location, or allowing for price discrimination with public dictation of the location of the facilities.

Given that in most of our work's appplication areas of the population is concentrated in urban areas, we base our analysis on the framework developed by Hwang and Mai [4]. Additionally, given that our goal is to analyse the effect of different policies on universal access rather than on total output, we depart from the assumption of elastic demand functions considering instead that demand is completely inelastic, up to certain reservation value, to the price of the good. As with this assumption each individual consumes either zero or one unit of the good, universal access takes place when the price charged to each agent, plus the transportation cost incurred, is less than the reservation value, which garantees the consumption by each customer.

Our results predict that the allowance for price discrimination ensures universal access more often. When this happens, the facilities are located in the most populated cities, and the inhabitants from other places are compensated for their journeys through a lower price. Full provision is more likely when the commuting population is more numerous, is located closer to the source and transport is cheaper. Public regulation of location tends to induce intermediate locations or positive consumer surplus but it does not improve the likelihood of full coverage.

# Some heuristic methods for the $p$-median problem with maximum distance constraints

Adrián Esteban Pérez[*],[1] Jesús Sáez-Aguado[2]

[1]*Universidad de Valladolid, Valladolid, Spain,*  adrianesteban@live.com

[2]*Departamento de Estadística e Investigación Operativa, Universidad de Valladolid, Valladolid, Spain,*  jsaez@eio.uva.es

**Abstract.** In this work we study the $p$-median problem with maximum distance constraints (**PMPDC**) which is a variant of the classical $p$-median problem (**PMP**). (**PMPDC**) appeared first time in [3] and it is a problem of interest in facility location. To our knowledge, the last study about heuristic methods for (**PMPDC**) is [1] based on Lagrangian relaxation. First of all, we provide some different formulations for (**PMPDC**). Note that (**PMP**) is a NP-hard problem, so adding the maximum distances constraints does not modify this complexity, but the problem is computationally more difficult. A first formulation is to consider the (**PMP**) with the following simple approach, based on modify the distance matrix:

$$d'_{ij} = \begin{cases} d_{ij}, & \text{if } d_{ij} \leq s_i \\ M, & \text{if } d_{ij} > s_i \end{cases}$$

where $d_{ij}$ is the distance between demand point $i$ and facility site $j$, $s_i$ is the the maximum distance limit between a demand point $i$ and any facility site and $M$ is a big value.

So we can transform (**PMPDC**) in a (**PMP**) with distance matrix modified. In a first look, (**PMPDC**) can be seen like a (**PMP**) but this formulation has a big problem: the heuristic methods for (**PMP**) frequently provide infeasible solutions and the quality of solutions depends of the value of big-$M$ ([1]).

---

[*]Corresponding author

We give other formulation, based on adding to (**PMP**) the maximum distance constraints and without modifying the distance matrix. Different heuristic procedures for the (**PMPDC**) problem are developed. First, a Lagrangian relaxation algorithm which differs from the existing in [1] is developed. Second, we apply a new approach, based on the GRASP methodology addapted to (**PMPDC**) from (**PMP**) [2].

In addition, we study in depth the relation between the feasibility of (**PMPDC**) and the parameters $p$ and maximum distance limits providing an analytic-geometric characterization.

Finally, in order to compare the different methods, two data sets have been used. The first set contains data from several real problems of medical assistance in Castilla & León, in Spain. The second data set contains some problems which are randomly generated (with bigger sizes than the first data set). We solve them very efficiently and we compare the obtained results with the exact solution computed with XPRESS.

**Keywords:** Facility location, $p$-median problem, Lagrangian relaxation, GRASP.

# On location and vessel fleet composition for offshore wind farm maintenance[*]

Alejandro Gutiérrez Alcoba,[1] Eligius M.T. Hendrix,[1] Gloria Ortega,[2] Dag Haugland,[3] Elin E. Halvorsen-Weare [4]

[1]*Computer Architecture, Universidad de Málaga*   agutierrez@ac.uma.es;eligius@uma.es

[2]*Informatics, Almería University,*   gloriaortega@ual.es

[3]*Department of Informatics, Bergen University,*   dag.haugland@uib.no

[4]*Department of Maritime Transport Systems, MARINTEK, Norway,* elin.halvorsen-weare@marintek.sintef.no

Maintenance provides a large part of the cost of an offshore wind farm. Several models have been presented in literature to optimize the fleet composition of the required vessels. A drawback such models is that they are based on perfect information on weather and incidences to schedule for the coming year. Our research question is what will happen to the fleet composition if the practical scheduling is simulated by using heuristics.

## 1.       Maintenance of offshore wind farms

The offshore wind energy industry is expected to continue its growth tendency in the near future. The European Wind Energy Association expects in its Central Scenario by 2030 a total installed capacity of 66 GW of offshore wind in the UE [1]. Offshore wind farms (OWFs) are large scale infrastructures, requiring large fleets able to perform operations and maintenance (O&M) activities on the installed turbines. The O&M cause a large part of the costs of running an OWF installation up to one third of the OWF costs, see [5]. Moreover, the fleet makes the installations depend on non-renewable energy resources. Therefore, optimising the efficiency of the resources used for the O&M activities of an OWF becomes extremely important in order to make them economically viable and to reduce CO2 emissions.

---

[*]Alejandro Gutierrez-Alcoba is a fellow of the Spanish FPI programme, granted by the Ministry of Economy, Industry and Competitiveness. This paper has been supported by The Spanish Ministry (TIN2015-66680) and Seneca Foundation (19241/PI/14) of the Murcia region, in part financed by the European Regional Development Fund (ERDF).

Recent deterministic and stochastic model formulations for vessel composition and maintenance optimization can be found in [2] and [3]. A recent literature review on DSS for OWF's is given by [4].

This basis of our investigation is a scenario based MILP model which like the models in [3] and [6] decide on the fleet composition. All these models evaluate the value of the vessel composition and base selection based on scheduling with perfect information; the weather conditions and breakdowns happening during a scenario of a year is known beforehand. Such a procedure underestimates the maintenance costs for a practical situation.

The research question is whether the composition may be affected a lot when maintenance scheduling is done in a more practical heuristic way given the practical information available.

# The mobile facility location problem

Mercedes Landete[1]

[1]*Universidad Miguel Hernández, 03202, Elche, Spain,*   landete@umh.es

The Mobile Facility Location Problem (MFLP) is the problem of re-locating a set of existing facilities and re-allocating all the customers so that the total cost of the movements is minimized. The re-location of facilities in a stochastic network which minimizes the expected travel time have been widely studied in the literature. However, the stochastic approach and the MFLP differ in two regards: firstly, the MFLP considers the initial location of the facilities, and thus the initial allocation of customers, as an input; secondly the objective function is the total movement cost instead of an expected total cost. The MFLP was introduced in [1]. Later, Halper et al. [2] presented a local heuristic and Raghavan and Sahin [3] considered the extension with capacitated facilities.

This work introduces two different set packing formulations for the problem. One formulation is for the case in which re-locating costs and re-allocating costs are proportional to distances and the another is for the general costs case. Valid inequalities and optimality conditions for each of the formulations are introduced. Computational results show the performance of both formulations as well as the performance of the different families of valid inequalities.

# Some criteria for locating sensors in a wind turbine blade

M.Cruz López-de-los-Mozos,[1] Juan A. Mesa,[2] Diego Ruiz-Hernández, [3] and Carlos Q. Gómez-Muñoz,[4]

[1]*Department of Applied Mathematics I, University of Seville, Spain,*   mclopez@us.es

[2]*Department of Applied Mathematics II, University of Seville, Spain,*   jmesa@us.es

[3]*Department of Quantitative Methods, University College for Financial Studies, Madrid, Spain,*   d.ruiz@cunef.edu

[4]*Department of Business Management, University of Castilla-La Mancha, Ciudad Real, Spain,*   carlosquiterio.gomez@uclm.es

Using acoustic sensors for detecting the location of a breakage on a wind turbine blade reduces the associated maintenance cost. The triangulation method developed in [1] is based on strategically placing three acoustic sensors in the surface of the blade section to approach the location of a randomly generated crack. Assuming the hypothesis made in [1], among them approximating the section blade by a planar surface (a rectangle), the question of how locating the three sensors is closely related with the accuracy of the approximation.

In this work we explore several criteria for locating acoustic sensors in a section of the wind turbine blade in order to apply the triangulation method to detect a breakdown in the surface of the blade.

Taking into account the experimental results obtained from the simulation of the method in the laboratory, we have considered several problems.

Let $\Omega$ denote the rectangle, and let $\Sigma = \{S_i, i = 1, \ldots, p\} \subset \Omega$ denote a set of $p$ sensors. We have considered a demand continuously distributed over $\Omega$, with uniform distribution in the first two cases.

1. $p$-**min-max-max criterion**. For $p = 3$, the problem is to find $\Sigma \subset \Omega$ minimizing the maximum distance from the point in $\Sigma$ furthest from

all points in $\Omega$:

$$\min_{\Sigma \subset \Omega,\ |\Sigma|=3} \tau(\Sigma) = \min_{\Sigma \subset \Omega,\ |\Sigma|=3}\ \max_{S \in \Sigma} \left\{ \max_{X \in \Omega} d(S, X) \right\}$$

$$\text{s.t.} \quad \min_{S_i, S_j \in \Sigma,\ S_i \neq S_j} d(S_i, S_j) \geq \delta > 0$$

2. **Maximum area criterion with sensors range threshold** Although in the laboratory the sensors cover the overall blade section, there are several blade sizes. In this case we have considered a sensors range threshold $\Delta > 0$, meaning that for $X \in \Omega$, if $d(X, S_i) > \Delta$ then $S_i$ does not receive the acoustic signal from $X$. Let $\mathcal{A}(conv(\Sigma))$ be the area of the convex hull of $\Sigma$. The second problem is

$$\max_{\Sigma \subset \Omega, |\Sigma|=3} \quad \mathcal{A}(conv(\Sigma))$$

$$\text{s.t.} \quad d(S, X) \leq \Delta,\ \forall X \in \Omega,\ \forall S \in \Sigma$$

3. **Maximum area criterion with regional fault distribution** The geometrical form make the edges of the wind turbine blade (trailing and leading edges) more susceptible to damage, in particular the trailing edge. Thus, we have considered three zones in $\Omega$, with different breaking probability. Let $\Omega_i, i = 1, \ldots, 3$ be three consecutive rectangles in $\Omega$ such that a side of $\Omega_1$ coincides with the trailing edge, and a side of $\Omega_3$ coincides with the leading edge Let $0 < w_1 \leq \min\{w_2, w_3\}$, with $W = \sum_{i=1}^{3} w_i$, and let $w_i/W$ be the probability of a fault in $\Omega_i$, $i = 1, \ldots, 3$. The third problem is

$$\max_{\Sigma \subset \Omega, |\Sigma|=3} F(\Sigma) := \frac{1}{W} \sum_{i=1}^{3} w_i \, \mathcal{A}(\Omega_i \cap conv(\Sigma))$$

The second phase of this work (still open) consists in incorporating these elements in a cooperative cover model, as the one proposed in [2]

# Tree of hubs location problem with upgrading [*]

Alfredo Marín[1]

[1]*Departamento de E. e Investigación Operativa, Universidad de Murcia,*  amarin@um.es

The Tree of Hubs Location Problem (THLP) was introduced for the first time in [2]. Since then, it has received much attention in the specialized literature. The THLP is a single-allocation hub location problem where $p$ hubs have to be located on a network and connected by means of a (nondirected) tree. Then each non-hub node must be connected (allocated) to a hub and all the flow between nodes must use these connections to circulate, i.e., excepting the arcs that connect each non-hub node with its allocated hub, the arcs that route the flows must be links connecting hubs. There is a per unit transportation cost associated with each arc. The objective is to minimize the operation costs of the system.

On the other hand, many variants of the Minimum Spanning Tree Problem are being explored, among them the Spanning Tree Problem with upgrading (STPU), see e.g. [1]. Here the cost (length, weight) associated to each edge of the graph can be reduced in the first instance by upgrading one of its extremes (with an associated cost), and it can be reduced even more by upgrading both extremes. In this way, every time a cost is paid to upgrade a node, all edges inciding in this node benefit from this upgrading.

We introduce here the THLP with upgrading (THLPU), a mixture of these two problems. In addition to locate the hubs, to determine the tree connecting hubs and to allocate non-hub nodes to hubs, a decision has to be taken about which of the hubs will be upgraded, taking into account that there is a budget to be invested in the upgrading. Then we formulate the problem as a Mixed Integer Linear Programming Problem, trying to get

---

[*]Research supported by Ministerio de Economía y Competitividad, project MTM2015-65915-R, Fundación Séneca, project 19320/PI/14, and Fundación BBVA, project "Cost-sensitive classification. A mathematical optimization approach" (COSECLA)

a tight formulation, and we generate several families of valid inequalities. A preliminary computational study is also presented.

# A stochastic multi-period covering model

Alfredo Marín,[1] Luisa I. Martínez-Merino,[2] Antonio M. Rodríguez-Chía,[3] Francisco Saldanha-da-Gama,[4]

[1]*Departamento de Estadística e Investigación Operativa, Facultad de Matemáticas, Universidad de Murcia, Murcia, Spain*,   amarin@um.es

[2]*Departamento de Estadística e Investigación Operativa, Universidad de Cádiz, Cádiz, Spain*,   luisa.martinez@uca.es

[3]*Departamento de Estadística e Investigación Operativa, Universidad de Cádiz, Cádiz, Spain*,   antonio.rodriguezchia@uca.es

[4]*Departamento de Estatística e Investigação Operacional/ Aplicações Fundamentais e Investigação Operacional, Faculdade de Ciências da Universidade de Lisboa, Portugal*,   faconceicao@fc.ul.pt

This work focuses on a general covering location problem, denoted as GSMC, which includes stochastic and multi-period features. It generalizes most of the covering models in the literature. In the GSMC, a planning horizon divided in several time periods is considered and uncertainty about the demand for coverage is also taken into account. Concretely, given a set of potential locations for facilities and a set of demand points, the purpose is to decide which facilities must be operating in each time period to minimize the total expected cost satisfying some coverage constraints. A formulation for this model is proposed and analyzed. In addition, a Lagrangian relaxation based heuristic is used not only to obtain lower bounds on the solutions, but also to find good feasible solutions for the model.

## 1.    Introduction to the problem

Two main classic covering models can be found in literature: the set covering location problem (SCP) proposed in [4], and the maximal covering location problem (MCLP) introduced in [1]. The objective of the former problem is to minimize the cost of installed facilities restricting that all demand points must be covered. The latter consists of maximizing the covered demand limiting the number of operating facilities. These two main models and some others related with covering are generalized in [2]. This general model is extended in the present work adding stochastic and multi-period features.

In the GSMC a finite planning horizon divided in a set of time periods ($T$) is considered. Besides, a set of potential locations for facilities ($I$) and a set of demand points ($J$) is given. In each potential location $i \in I$, a maximum of $e_i$ facilities can be operating and only a total of $p_t$ facilities can be active in each time period $t \in T$. When a facility is installed or closed in a certain time period a cost must be paid. Similarly, an operating cost for the activity of a facility in a time period is considered.

The GSMC model also assumes that it exists uncertainty associated with the minimum threshold for coverage of each demand point and with the coverage capability of each facility. Moreover, a profit related with the number of facilities covering a demand point above its minimum threshold, and a penalty associated with the coverage shortage are modeled. These profits/penalties are also uncertain. It is considered that the uncertainty can be explained by a finite set of scenarios with some previously known probabilities.

Given the previous framework, GSMC model aims to decide which facilities must be installed or closed in each time period to minimize the total expected cost satisfying some coverage constraints. As a result, GSMC generalizes not only the models appearing in [2], but also some other covering models that include some time-dependent parameters, see [3].

In addition to the model analysis, a Lagrangian relaxation based heuristic which provides lower bounds and good feasible solutions is developed. Some preliminary computational results were performed to see the importance of this Lagrangian heuristic.

# Supply chain complexity and the network design : Location does matter!

Mozart B.C. Menezes, [2,3] Diego Ruiz-Hernández [1,3]

[1] *University College for Financial Studies, Department of Quantitative Methods, Leonardo Prieto Castro 2, 28040, Madrid, Spain*   d.ruiz@cunef.edu

[2] *Kedge Business School, Operations Management and Information Systems Department, KEDGE Business School Bordeaux, 680 Cours de la Libération, 33405, Cedex, France*

[3] *Supply Chain and Complexity Lab, KEDGE Business School*

Facility location problems are well known problems in the field of combinatorial optimization, where -broadly speaking- the objective is typically to locate a collection of facilities aimig at minimising the cost incurred in serving the customer base. There are several papers dealing with facility location algorithm complexity starting by showing that most belong to the class of NP-hard problems (e.g., see [2]). Additionally, some very sophisticated results take advantage of structural properties of some location problems and present a bound on the performance of a greedy algorithm [1]. In this paper we focus on another type of complexity. Our study brings to the community of facility location the concept of operations complexity, where the objective is to measure the amount of information those managing a supply network have to deal with. THis new concept opens up a new research line within the field. When determining the location of a facility one should aim not only at reducing operational costs but also at keeping (operational) complexity in what we call a *complexity comfort range*, in which tactical and operational decisions are at their bests. Preliminary (empirical) results suggest that ignoring complexity issues may hurt that same bottom line that the locational problem is trying to improve.

# Heuristics for the stochastic uncapacitated $r$-allocation $p$-hub median problem

Juanjo Peiró[1], Ángel Corberán[1], Rafael Martí[1] and
Francisco Sandanha-da-Gama[2]

[1]*Departament d'Estadística i Investigació Operativa. Universitat de València, Spain.*
juanjo.peiro@uv.es, angel.corberan@uv.es, rafael.marti@uv.es

[2]*Departamento de Estatística e Investigação Operacional.*
*Centro de Matemática Aplicações Fundamentais e Investigação Operacional,*
*Faculdade de Ciências, Universidade de Lisboa, Portugal*
fsgama@ciencias.ulisboa.pt

In this work we study a class of hub median problems that has been referred to in the literature as the $r$-allocation $p$-hub median problem. We consider an existing modeling framework and extend it by including several features that include fixed allocation costs, non-stop services between terminals and stochasticity in the traffic and transportation costs. For the situation in which the support of the underlying random vector is finite we propose a heuristic algorithm for finding high-quality feasible solutions.

## 1.     Introduction

The starting point for your study is the so-called uncapacitated $r$-allocation $p$-hub median problem (U$r$A$p$HMP) introduced by Yaman [1] and studied by other authors such as Peiró et al. [2] and Martí et al. [3]. In this problem, a set of nodes $V$ is given such that some traffic $t_{ij}$ must be routed between many pairs of nodes $(i, j) \in V \times V$. The goal is to select a set $H \subseteq V$ with $|H| = p$ for installing hubs. In this model, the hub network is assumed to be complete and all the traffic must be routed via at least one hub (direct shipments are not possible). Each node can be allocated to at most $r$ hubs, where $r$ is exogenously defined. The goal is to minimize the total trans-portation cost. The unitary transportation costs are assumed to satisfy the triangle inequality.

# 2. Extensions to the uncapacitated $r$A$p$HMP

We extend the above mentioned model in several directions namely, by considering: (i) fixed allocation costs, (ii) the possibility of having direct transportation between terminals, and (iii) uncertainty in transportation costs and traffics. We call the extended problem, the stochastic uncapacitated $r$-allocation $p$-hub median problem with non-stop services—Stochastic U$r$A$p$HMP-NSS.

This new setting is much general since it captures several particular cases/problems of interest such as the stochastic single allocation $p$-hub median problem, the stochastic multiple allocation $p$-hub median problem, and each of those problems combined with the possibility of including allocation costs and/or non-stop services. Hence, we are not studying a particular problem of interest but a broader setting that captures several particular problems that may be of interest.

We assume that uncertainty can be described probabilistically using a joint distribution function know in advance (e.g., instance estimated using historical data). This makes intuitive the use of a stochastic programming modeling framework: in the first stage we consider the network design decisions (location and allocation decisions); in the second stage (after uncertainty is disclosed) we consider the transportation decisions.

Assuming that the support of the underlying random vector is finite, we can go further in terms of mathematical modeling and derive a compact formulation for the deterministic equivalent of the stochastic problem developed. Unfortunately, even for very small instances of the problem, the model becomes too large, which prevents the use of a general-purpose solver for tackling it. This motivates the development of an approximate procedure for finding feasible solutions to the problem.

# Profiling the inherent complexity of different facility location strategies

Jesús María Pinar-Pérez,[1], Diego Ruiz-Hernández, [1,3] Mozart Menezes,[2,3]

[1]*University College for Financial Studies, Department of Quantitative Methods, Leonardo Prieto Castro 2, 28040, Madrid, Spain*   jesusmaria.pinar@cunef.edu

[2]*Kedge Business School, Operations Management and Information Systems Department, KEDGE Business School Bordeaux, 680 Cours de la Lib'eration, 33405, Cedex, France*

[3]*Supply Chain and Complexity Lab, KEDGE Business School*

Facility location problems are well known combinatorial problems where the objective is to minimize the cost incurred to serve customers from a set of facilities. Our aim is to bring to the field of facility location the concept of operations complexity, opening up a new research line. The main objective is to create awareness about the need of considering complexity issues and its impact on profitability, rather than only a cost/profit perspective, when deciding the location and size of a distribution network.

In our preliminary numerical assessment, we consider both randomly generated and real life networks, taking into consideration their temporal evolution, using proxies for organic growth, mergers and acquisitions and optimised design. We also evaluate the effect of different opimisation strategies and provide insights for further development and discussion.

# Approval voting problem under the $k$-centrum criterion

Diego Ponce,[1] Justo Puerto,[2] Federica Ricca,[3] and Andrea Scozzari[4]

[1] *Universidad de Sevilla, Dep. de Estadística e Investigación Operativa, IMUS,* dponce@us.es

[2] *Universidad de Sevilla, Dep. de Estadística e Investigación Operativa, IMUS,* puerto@us.es

[3] *Sapienza Universitá di Roma, Dip. MEMOTEF* federica.ricca@uniroma1.it

[4] *Universitá degli Studi Niccoló Cusano, Facoltá di Economia* andrea.scozzari@unicusano.it

In this work we model the approval voting problem as a mixed integer linear program. Different formulations for the Minisum, Minimax and $k$-centrum objective functions have been developed. The usefulness of these new approach to solve this problem is evaluated with a computational study.

## 1.     Introduction

Consider a set of $n$ voters and a set of $m$ candidates. Let $P$ be a $n \times m$ matrix representing the set of $n$ voters profiles, that is, $P_i$, $i = 1, \ldots, n$, is a boolean vector of length $m$ that specifies each voter's preference on each candidate. For instance, $p_{ij} = 1$ means that voter $i$ approves candidate $j$; $p_{ij} = 0$ otherwise. The problem consists of selecting a committee, that is, a (sub)set of candidates in order to minimize a given objective function.

Given a committee $x$ (a boolean vector $x$ of length $m$, $x_j = 1$ if candidate $j$ belongs to the committee; $x_j = 0$ otherwise), the Hamming distance is used to evaluate the distance between a committee and each profile $P_i$, $i = 1, \ldots, n$. Let $x$ be a committee, the Hamming distance $d_i(x)$ between the profile $P_i$ of voter $i$ to $x$ is $d_i(x) = \sum_{j=1}^{m} |p_{ij} - x_j|$.

The two criteria commonly used for electing a committee are:

1. Minisum criterion: Elect a committee minimizing the sum of Hamming distances to the profiles.

2. Minimax criterion [2]: Elect a committee minimizing the maximum of Hamming distances to the profiles.

From a computational point of view, the Minisum problem is polynomially solvable [1]. The aim of this work is to consider criteria that are *between* the Minisum and the Minimax ones. This consists of a family of functions, parameterized by a vector $W$ of length $n$, mapping a vector of scores $H$ to an aggregated score.

## 2.     The $k$-centrum voting problem

The family of Approval Voting rules parameterized by a vector $W$ that we will consider here is

$$W(k) = (\underbrace{1, \ldots, 1}_{k}, 0, \ldots, 0)$$

where $k$ is the number of ones. For example, if $k = n$ we have the Minisum criterion, if $k = 1$ we have the Minimax criterion. We want to study the general case when $1 < k < n$.

More specifically, given $1 < k < n$, the problem turns out to be: elect a committee that minimizes the sum of the $k$-largest weights ($k$-centrum). Let $\sigma(x)$ be an ordering function such that $d_{\sigma_1}(x) \geq d_{\sigma_2}(x) \geq \ldots \geq d_{\sigma_n}(x)$. The problem can be formulated as follows:

$$\min_x \sum_{h=1}^{k} d_{\sigma_h}(x) : x \in \{0, 1\}^m. \tag{1}$$

In this work, we propose several formulations and methodologies which are experimentally compared in a data base from the literature.

# The ordered median tree of hubs location problem

Miguel A. Pozo,[1] Justo Puerto,[2] and Antonio M. Rodríguez-Chía,[3]

[1]*Universidad de Cádiz, Spain.,*   miguelpozo@us.es

[2]*Universidad de Sevilla, Spain.,*   puerto@us.es

[3]*Universidad de Cádiz, Spain.,*   antonio.rodriguezchia@uca.es

Hub-and-spoke models have a great importance in transportation and telecommunication systems in which several origin-destination points exchange flows. In such models the key feature is to connect each pair via specific subsets of links to consolidate, and distribute the flows in order to reduce costs based on the economy of scale of intermediate connections. Therefore, Hub Location Problems integrate two level of decisions: location of facilities (hubs) to consolidate deliveries and network design to determine the routes that different origin-destination pairs follow to improve performance.

The standard model of hub-and-spoke networks assume that inter-hub connections between an origin-destination pair can be routed through one or at most two hubs. However, it has been observed by several authors that in many applications the backbone network is not fully interconnected [1,2] or it can even be not necessarily connected.It is of special interest the case where the underlying interconnection network is connected by means of a tree. Such problem is called the Tree of Hubs Location Problem and was introduced by Contreras et al ( [3,4]).

Recently, another feature, namely weighted averaging objective functions, has also been incorporated to the analysis of Hub Locations Problems [5, 6]. It has been recognized as a powerful tool from a modeling point of view because its use allows to distinguish the roles played by the different entities participating in a hub-and-spoke network inducing new type of distribution patterns. Each one of the components of any origin-destination delivery path gives rise to a cost that is weighted by different compensation factors depending on the role of the entity that supports the cost. This adds a "sorting"-problem to the underlying hub location problem. The objective is to minimize the total transportation cost of the flows between each origin-destination pair after applying rank dependent compensation factors on the transportation costs.

In this paper, we propose the Ordered Median Tree of Hub Location Problem (OMTHLP). The OMTHLP is a single allocation hub location problem where $p$ hubs must be placed on a network and connected by a non-directed tree. Each non-hub node is assigned to a single hub and all the flow between origin-destination pairs must circulate using the links connecting the hubs. The objective is to minimize the sum of the ordered weighted averaged assignment costs plus the sum of the circulating flow costs. We will present different MILP mathematical formulations for the OMTHLP based on the properties of the Minimum Spanning Tree Problem and the Ordered Median optimization. We establish a theoretical and empirical comparison between these new formulations and we also provide reinforcements that together with a proper formulation are able to solve medium size instances on general graphs.

# Location theory and some physical principles in a nutshell

Justo Puerto,[1]

[1]*IMUS. Universidad de Sevilla, Spain.,*   puerto@us.es

Location Theory is an appealing field of Operations Research that shares many links with optimization. The most important problems in this area can be formulated as mathematical programming problems in different framework spaces: continuous, networks or discrete [9]. Lots of insights are gained on the original problems analyzing their mathematical programs counterparts and reciprocally, many results can be inherited from the structure of the original problems; what helps in solving the models.

This presentation would like to focus on a different aspect that is not so well-known: the relationship between classical principles in Physics and some models and solution methods applied in standard location problems.

We will revisit some well-known laws as the equilibrium of forces, symmetry, maximum entropy, law of Snell, law of gravity or the optimal mass transport theory [6]. The goal will be to link them to some problems in the field of Location analysis [9].

In this talk, we show how the equilibrium of forces can be used to derive algorithms to solve continuous (single and multiple) facility location [1, 2, 4], symmetry is the basis of the ordered median problem [10], maximum entropy can explain obnoxious facility location models, the law of Gravity can be used to determine market shared areas or to define territorial units [5,7], the Snell's law is applicable to modeling different transport modes in location or transportation problems [3] or how the optimal mass transport theory can be used to compute optimal territory design [8]. We will explain the connections, interpret the results in the *jargon* of location analysis and show some applications.

# The periodic vehicle routing problem with driver consistency[*]

Inmaculada Rodríguez-Martín,[1] Juan-José Salazar-González,[1] and Hande Yaman[2]

[1] *DMEIO, Facultad de Ciencias, Universidad de La Laguna, Tenerife, Spain*  irguez@ull.es, jjsalaza@ull.es

[2] *Dpt. of Industrial Engineering, Bilkent University, Ankara, Turkey*  hyaman@bilkent.edu.tr

The Periodic Vehicle Routing Problem is a generalization of the classical VRP in which routes are determined for a planning horizon of several days. Each customer has an associated set of allowable visit schedules, and the objective of the problem is to design a set of minimum cost routes that give service to all the customers respecting their visit requirements. In this paper we study a variant of this problem in which we impose that each customer should be served by the same vehicle/driver at all visits. We call this problem the Periodic Vehicle Routing Problem with Driver Consistency (PVRP-DC). We present different integer linear programming formulations for the problem and derive several families of valid inequalities. We solve it using an exact branch-and-cut algorithm, and show computational results on a wide range of randomly generated instances.

--------------------
[*]This work has been partially supported by the Spanish research project MTM2015-63680-R (MINECO/FEDER)

# The effect of products' short lifecycle on network design

Diego Ruiz-Hernández,[1,3] Mozart B.C. Menezes,[2,3] and Oihab Allal-Cheriff[2]

[1]*University College for Financial Studies, Department of Quantitative Methods, Leonardo Prieto Castro 2, 28040, Madrid, Spain* d.ruiz@cunef.edu

[2]*Kedge Business School, Operations Management and Information Systems Department, KEDGE Business School Bordeaux, 680 Cours de la Lib'eration, 33405, Cedex, France*

[3]*Supply Chain and Complexity Lab, KEDGE*

In this work we address the problem of designing a distribution network for new/seasonal products. The complexity of this problem becomes magnified because the location/capacity decisions are made long before the product is released to the market and, thus, knowledge about its demand is limited. Moreover, in general new (or seasonal) products show very short life-cycles, making location and capacity one-shot decisions. An example of that could be the production of a completely new model of car, where the assembly facilities are designed with a fixed number of lines and changes to the initial decision are both costly and not very simple, if possible at all. Another example, could be the case of temporary facilities for humanitarian aid, where adapting existing buildings for housing a supply center requires some important investment (including security of the premises) and it is a one go decision. In this case, the need for the facility is short lived while the real demand is highly uncertain. The relevance of location and capacity decisions is better appreciated by considering that about 80% of the supply chain costs are locked-in once the facilities' location and capacity are fixed [6].

Assuming that production can either be fully manufactured in-house or partially outsourced, we imply that, when outsourcing, the firm has certain flexibility with respect to the quantities ordered. The resulting trade-off between over- and under-capacity is hereby exploited for defining the capacity of each facility which, consequently, has an impact on its location.

The model that has traditionally been called-for when dealing with capacity issues is the Newsvendor model. Regarding location, the deterministic variant of the capacitated facility location problems has been thoroughly addressed in literature, see for example [1, 3]. However, stochasticity has only recently been included in capacitated frameworks. See, for example, [4], [2], or [5].

To our knowledge, this is the first time that the problem of simultaneously locating facilities and determining their capacities is addressed with full consideration of demand's stochasticity. We show that for the single facility case, the expected profit of the strategic problem is non-decreasing and concave in the facility capacity, resulting in a uniquely determined optimal capacity. We further show that when the facility's location is fixed, the problem of choosing the capacity becomes a variation of the classical Newsvendor model. A critical-ratio based heuristic is proposed for the multi-facility case. We finally provide an illustrative example.

# Facilities delocation in the retail sector

María Sierra-Paradinas,[1] Antonio Alonso-Ayuso,[2] and J. Francisco Rodríguez-Calo [3]

[1]*Escuela Técnica Superior de Ingeniería Informática, Universidad Rey Juan Carlos, Madrid, Spain,* maria.sierrap@urjc.es

[2]*Escuela Técnica Superior de Ingeniería Informática, Universidad Rey Juan Carlos, Madrid, Spain,* antonio.alonso@urjc.es

[3]*Repsol S.A., Madrid, Spain,* jfrodriguezc@repsol.com

A facilities delocation problem in the retail sector is addressed by proposing a mixed integer programming model. The aim of the problem is to decide the cease of existing facilities in order to optimize the income of the whole retail network.

## 1. Introduction

Facilities location models have been broadly studied in the literature. Nevertheless, the concept of delocation has been introduced recently. In [1] delocation is defined as the operation ceased of existing facilities.

As shown in [2] several delocation models have been presented in both the private and the public sector. In particular, [2] presents two models to reduce the number of facilities in a given area with firm competition and without it. In the first case their aim is to reduce the quantity of facilities to a fixed number, minimizing the impact on demand loss to competitors. In the second case, a measure of the decline of the service is minimized.

Along the same line, in [1] a model to downsize a firm's existing distribution network with known supplier locations is presented. The firm seeks the closure of a fixed number of the supplier nodes. The model proposed contemplates that all demand nodes must be served from their respective supplier except if the existing supplier is eliminated.

More recently in [3] a model for resizing a bank network is presented. Seeking to maintain a constant service level, the objective is to decide which branches should stay opened within a colletion of possibly redundant ones.

# 2.    Definition of the problem

Delocation models have been addressed in the literature due to several reasons. In our case a franchise chain wants to optimize the operation of his network of stores. The decision to make is whether the existing stores should cease operations or change the operating mode.

The stores can be operated by the franchise chain itself or by an external dealer. Each operating mode has a different repercusion in the final income of the stores due to the client's behaviour, when this behaviour is given by their tendency to abandon. In case of the closure of a store the clients with tendency to abandon will leave the entire franchise chain and will not report any benefits. If the store's operating mode is changed, the final prices and final income change as well, but no repercusion in the clients is noticed.

Capacity constraints are imposed in the number of stores that should stay opened and cease operations cost, client behavoiur and final prices depending on the operating mode are known.

Due to business demands refering to the behaviour of the clients, a non-linear constraint appears in the definition of the model. Fortet's innequalities are used in order to linearize the constraint and therefore obtain an integer linear programming model.

Because of the size of the network, border constraints have been imposed in order to get results in a reasonable computational time and model optimization is done by introducing smart index sets in order to reduce the number of constraints and variables. The model has been written into AMPL and solved using CPLEX version 12.7.

# Author Index

VIII Work
Locat
Analysis
Proble

SPONSORS:

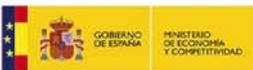
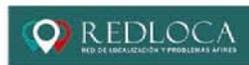
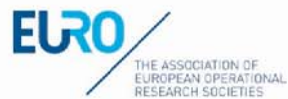
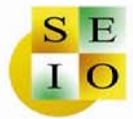

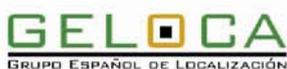
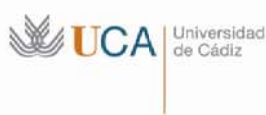
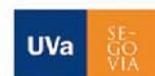